\documentclass[11pt,twoside]{article}
\usepackage{asp2010}

\resetcounters

\begin{document}

\title{The Sloan Digital Sky Survey Data Release 7 M Dwarf Spectroscopic Catalog}
\author{Andrew A. West$^{1}$, Dylan P. Morgan$^{1}$, John J. Bochanski$^{2,3}$, Jan Marie Andersen$^{1}$, Keaton J. Bell$^{4}$, Adam F. Kowalski$^{4}$, James R. A. Davenport$^{4}$, Suzanne L. Hawley$^{4}$, Sarah J. Schmidt$^{4}$, David Bernat$^{5}$, Eric J. Hilton$^{4}$, Philip Muirhead$^{5}$, Kevin R. Covey$^{1,5}$, B{\'a}rbara Rojas-Ayala$^{5}$, Everett Schlawin$^{5}$, Mary Gooding$^{6}$, Kyle Schluns$^{1}$, Saurav Dhital$^{7}$, J. Sebastian Pineda$^{3}$, David O. Jones$^{1}$
\affil{$^1$Department of Astronomy, Boston University, 725 Commonwealth Avenue, Boston, MA 02215, USA, email: aawest@bu.edu, dpmorg@bu.edu, janmarie@bu.edu, kcovey@astro.cornell.edu, kschluns@bu.edu, jonesd@bu.edu}
\affil{$^2$Astronomy and Astrophysics Department, Pennsylvania
  State University, 525 Davey Laboratory, University Park, PA 16802, USA,  
email:jjb29@psu.edu}
\affil{$^3$Kavli Institute for Astrophysics and Space Research, Massachusetts Institute of Technology, Building 37, 77 Massachusetts Avenue, Cambridge, MA 02139, USA, email: jspineda@caltech.edu}
\affil{$^{4}$Astronomy Department, University of Washington,
   Box 351580, Seattle, WA  98195, USA, email: keatonbell@comcast.net, kowalski@astro.washington.edu, jrad@astro.washington.edu, slh@astro.washington.edu, sjschmidt@astro.washington.edu, hilton@astro.washington.edu}
\affil{$^5$ Department of Astronomy, Cornell University, 610 Space Sciences Building, Ithaca, NY 14853, USA, email: david.bernat@gmail.com, muirhead@astro.cornell.edu, babs@astro.cornell.edu, everett.schlawin@gmail.com}
\affil{$^6$ Wells College, Department of Mathematical and Physical Sciences, 170 Main Street, Aurora, NY 13026, USA, email: mgooding256@gmail.com}
\affil{$^7$ Physics and Astronomy Department, Vanderbilt University, 6301 Stevenson Center, Nashville, TN 37235, USA, email: saurav.dhital@vanderbilt.edu}}

\begin{abstract}
  We present a spectroscopic catalog of 70,841 visually inspected M
  dwarfs from the seventh data release (DR7) of the Sloan Digital Sky
  Survey (SDSS). For each spectrum, we provide measurements of the
  spectral type, a number of molecular bandheads, and the H$\alpha$,
  H$\beta$, H$\gamma$, H$\delta$ and Ca II K emission lines. In
  addition, we calculate the metallicity-sensitive parameter $\zeta$
  and 3D space motions for most of the stars in the sample. Our
  catalog is cross-matched to Two Micron All Sky Survey (2MASS)
  infrared data, and contains photometric distances for each
  star. Future studies will use these data to thoroughly examine
  magnetic activity and kinematics in late-type M dwarfs and examine
  the chemical and dynamical history of the local Milky Way.
\end{abstract}

\section{Description of the Catalog}

We provide a brief description of some of the measured quantities in
our Sloan Digital Sky Survey \citep[SDSS;][]{york00} Data Release 7
\citep[DR7;][]{dr7} value-added catalog below.  A much more thorough
description of the catalog selection and its bulk characteristics can
be found in \citet{west10}.  The catalog will eventually be available
on the Vizier\footnote{http://vizier.u-strasbg.fr/cgi-bin/VizieR}
site, but can also be obtained immediately by contacting the primary
author (AAW).

\begin{figure}[!ht]
 \plotone{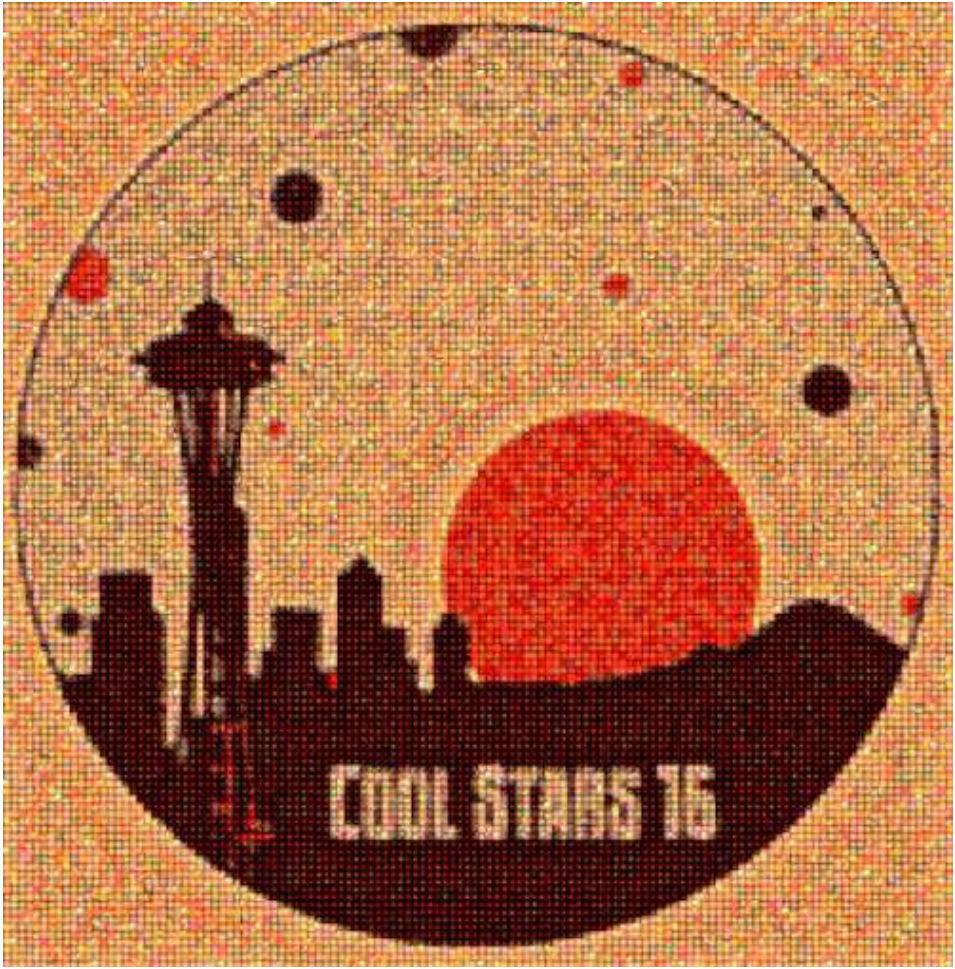}
 \caption{2$^{\prime\prime} \times 2^{\prime\prime}$ $gri$ color composite images of more than 3000 M dwarfs from the SDSS DR7 M dwarf catalog. The images have been mosaicked to reproduce the Cool Stars 16 logo (original logo credit: L. Walkowicz).} 
 \label{fig:bam}
\end{figure}

We visually inspected 116,161 M dwarf candidates (color selected from
the SDSS database) and manually assigned spectral types.  The sample
was divided among 17 individuals\footnote{The order of the co-authors
  was based on the number of spectra examined.} who used the manual
``eyecheck'' mode of the Hammer \citep[v.\ 1\_2\_5;][]{covey07} to
assign spectral types and remove non-M dwarf interlopers, resulting in
70,841 M dwarfs (see Figure \ref{fig:bam}).  We also matched our
catalog to the 2MASS point source catalog \citep{2mass}, matching only
to unique 2MASS counterparts within 5$^{\prime\prime}$ of the SDSS
position that do not fall within the boundaries of an extended source
({\sc gal\_contam} $=$0).

Radial velocities (RVs) were measured by cross-correlating each
spectrum with the appropriate \citet{bootem} M dwarf template.  This
method has been shown to produce uncertainties ranging from 7-10 km\
s$^{-1}$ \citep{bootem}.  All of the DR7 objects were cross-matched to
the USNO-B/SDSS proper motion catalog \citep{munn04,munn08},
identifying 39,151 M dwarfs with good proper motions.  Distances to each star were
calculated using the $M_r$ vs.\ $r-z$ color-magnitude relation given
in \citet{boo10}.  The proper motions and distances were combined with the RVs
to produce 3-dimensional space motions for the DR7 M dwarfs.

As part of our analysis, we measured a number of spectral lines and
molecular features in each M dwarf spectrum.  All of the spectral
measurements were made using the RV corrected spectra.  The TiO1,
TiO2, TiO3, TiO4, TiO5, TiO8, CaOH, CaH1, CaH2, and CaH3 molecular
bandhead indices and their formal uncertainties were measured using
the Hammer with the molecular bandheads as defined by \citet{pmsu1}
and \citet{gizis97}.  We also measured the chromospheric hydrogen
Balmer and Ca II lines that are associated with magnetic activity.  We
expanded the H$\alpha$ analysis of \citet{west04, west08} to include
H$\beta$, H$\gamma$, H$\delta$ and Ca II K (H$\epsilon$ and Ca II H
are blended in SDSS data and were not included in our sample).  All of
the line measurements were made by integrating over the specific line
region (8 \AA\ wide centered on the line) and subtracting off the mean
flux calculated from two adjacent continuum regions.  Equivalent
widths (EW) were computed for each line by dividing the integrated
line flux by the mean continuum value.

For all of the active stars in the sample \citep[see][for
definition of activity]{west10} we computed the ratio of luminosity in the emission
line as compared to the bolometric luminosity
(L$_{\rm{line}}$/L$_{\rm{bol}}$).  We followed the methods of
\citet{hall96}, \citet{walkowicz04} and \citet{westhawley08} who
derived $\chi$ factors for the Balmer and Ca II chromospheric lines as
a function of M dwarf spectral type.  The
L$_{\rm{line}}$/L$_{\rm{bol}}$ values were computed by multiplying the
EW of each active star by the appropriate $\chi$ value.  Formal
uncertainties were computed for each L$_{\rm{line}}$/L$_{\rm{bol}}$
value and are included in the final database.

We also computed the metal sensitive parameter $\zeta$, defined by
\citet{lepine07}, which uses a combination of the TiO5, CaH2, and CaH3
molecular band indices to separate the sample into different
metallicity classes.  This is similar to the
\citet{gizis97} classification system but was re-calibrated using wide
common proper motion pairs that were assumed to be at the same
metallicity.  Stars with solar metallicity ([Fe/H]=0) have $\zeta$
values $\sim$1 and stars with [Fe/H]=-1 have $\zeta$ $\sim$0.4
\citep{woolf09}.  Although there is considerable scatter in the [Fe/H]
versus $\zeta$ relation at high-metallicities, this parameter is very
useful for finding and classifying low-metallicity stars that are
likely members of the Galactic halo.

As with previous SDSS spectroscopic catalogs of low-mass stars, we
remind the community that these data do not represent a complete
sample and that the complicated SDSS spectral targeting introduces a
variety of selection effects.  However, our new sample covers a large
range of values for many of the physical attributes of the M dwarfs,
including parameters that are sensitive to activity, metallicity, and
Galactic motion, making accurate activity, kinematic, and chemical
analyses possible. In addition, because some of the derived quantities
are computed by automatic routines, values for a small percentage of
individual stars may be incorrect; this should not affect large
statistical results. Users are nevertheless cautioned to understand
the origin of specific data products before using them
indiscriminately.

\begin{thebibliography}{}
\expandafter\ifx\csname natexlab\endcsname\relax\def\natexlab#1{#1}\fi
\expandafter\ifx\csname url\endcsname\relax
  \def\url#1{\texttt{#1}}\fi
\expandafter\ifx\csname urlprefix\endcsname\relax\def\urlprefix{URL }\fi
\providecommand{\eprint}[2][]{\url{#2}}

\bibitem[{{Abazajian} et~al.(2009)}]{dr7}
{Abazajian}, K.~N., et~al. 2009, \apjs, 182, 543. \eprint{0812.0649}

\bibitem[{{Bochanski} et~al.(2010){Bochanski}, {Hawley}, {Covey}, {West},
  {Reid}, {Golimowski}, \& {Ivezi{\'c}}}]{boo10}
{Bochanski}, J.~J., {Hawley}, S.~L., {Covey}, K.~R., {West}, A.~A., {Reid},
  I.~N., {Golimowski}, D.~A., \& {Ivezi{\'c}}, {\v Z}. 2010, \aj, 139, 2679.
  \eprint{1004.4002}

\bibitem[{{Bochanski} et~al.(2007){Bochanski}, {West}, {Hawley}, \&
  {Covey}}]{bootem}
{Bochanski}, J.~J., {West}, A.~A., {Hawley}, S.~L., \& {Covey}, K.~R. 2007,
  \aj, 133, 531. \eprint{arXiv:astro-ph/0610639}

\bibitem[{{Covey} et~al.(2007)}]{covey07}
{Covey}, K.~R., et~al. 2007, \aj, 134, 2398. \eprint{0707.4473}

\bibitem[{{Cutri} et~al.(2003)}]{2mass}
{Cutri}, R.~M., et~al. 2003, {2MASS All Sky Catalog of point sources.}

\bibitem[{{Gizis}(1997)}]{gizis97}
{Gizis}, J.~E. 1997, \aj, 113, 806. \eprint{arXiv:astro-ph/9611222}

\bibitem[{{Hall}(1996)}]{hall96}
{Hall}, J.~C. 1996, \pasp, 108, 313

\bibitem[{{L{\'e}pine} et~al.(2007){L{\'e}pine}, {Rich}, \& {Shara}}]{lepine07}
{L{\'e}pine}, S., {Rich}, R.~M., \& {Shara}, M.~M. 2007, \apj, 669, 1235.
  \eprint{0707.2993}

\bibitem[{{Munn} et~al.(2004){Munn}, {Monet}, {Levine}, {Canzian}, {Pier},
  {Harris}, {Lupton}, {Ivezi{\'c}}, {Hindsley}, {Hennessy}, {Schneider}, \&
  {Brinkmann}}]{munn04}
{Munn}, J.~A., {Monet}, D.~G., {Levine}, S.~E., {Canzian}, B., {Pier}, J.~R.,
  {Harris}, H.~C., {Lupton}, R.~H., {Ivezi{\'c}}, {\v Z}., {Hindsley}, R.~B.,
  {Hennessy}, G.~S., {Schneider}, D.~P., \& {Brinkmann}, J. 2004, \aj, 127,
  3034

\bibitem[{{Munn} et~al.(2008){Munn}, {Monet}, {Levine}, {Canzian}, {Pier},
  {Harris}, {Lupton}, {Ivezi{\'c}}, {Hindsley}, {Hennessy}, {Schneider}, \&
  {Brinkmann}}]{munn08}
--- 2008, \aj, 136, 895

\bibitem[{{Reid} et~al.(1995){Reid}, {Hawley}, \& {Gizis}}]{pmsu1}
{Reid}, I.~N., {Hawley}, S.~L., \& {Gizis}, J.~E. 1995, \aj, 110, 1838

\bibitem[{{Walkowicz} et~al.(2004){Walkowicz}, {Hawley}, \&
  {West}}]{walkowicz04}
{Walkowicz}, L.~M., {Hawley}, S.~L., \& {West}, A.~A. 2004, \pasp, 116, 1105.
  \eprint{arXiv:astro-ph/0410422}

\bibitem[{{West} \& {Hawley}(2008)}]{westhawley08}
{West}, A.~A., \& {Hawley}, S.~L. 2008, \pasp, 120, 1161. \eprint{0812.1221}

\bibitem[{{West} et~al.(2008){West}, {Hawley}, {Bochanski}, {Covey}, {Reid},
  {Dhital}, {Hilton}, \& {Masuda}}]{west08}
{West}, A.~A., {Hawley}, S.~L., {Bochanski}, J.~J., {Covey}, K.~R., {Reid},
  I.~N., {Dhital}, S., {Hilton}, E.~J., \& {Masuda}, M. 2008, \aj, 135, 785.
  \eprint{0712.1590}

\bibitem[{{West} et~al.(2004){West}, {Hawley}, {Walkowicz}, {Covey},
  {Silvestri}, {Raymond}, {Harris}, {Munn}, {McGehee}, {Ivezi{\'c}}, \&
  {Brinkmann}}]{west04}
{West}, A.~A., {Hawley}, S.~L., {Walkowicz}, L.~M., {Covey}, K.~R.,
  {Silvestri}, N.~M., {Raymond}, S.~N., {Harris}, H.~C., {Munn}, J.~A.,
  {McGehee}, P.~M., {Ivezi{\'c}}, {\v Z}., \& {Brinkmann}, J. 2004, \aj, 128,
  426. \eprint{arXiv:astro-ph/0403486}

\bibitem[{{West} et~al.(2010)}]{west10}
{West}, A.~A., et~al. 2010, \aj, submitted

\bibitem[{{Woolf} et~al.(2009){Woolf}, {L{\'e}pine}, \&
  {Wallerstein}}]{woolf09}
{Woolf}, V.~M., {L{\'e}pine}, S., \& {Wallerstein}, G. 2009, \pasp, 121, 117

\bibitem[{{York} et~al.(2000)}]{york00}
{York}, D.~G., et~al. 2000, \aj, 120, 1579. \eprint{arXiv:astro-ph/0006396}

\end{thebibliography}
\end{document}